\documentclass[
 reprint,twocolumn,
 amsmath,amssymb,
 aps,
]{revtex4}

\usepackage{epsfig,bm,epsf,graphics}
\usepackage{graphicx}
\usepackage{dcolumn}
\usepackage{bm}

\begin{document}


\title{Phase Transition and Surface Sublimation of a Mobile Potts Model} 



\author{A. Bailly-Reyre$^1$}
\email[E-mail: ]{aurelien.bailly-reyre@u-cergy.fr}
\author{H. T. Diep$^1$}
\email[E-mail: ]{diep@u-cergy.fr}

\author{M. Kaufman$^{1,2}$}
\email[E-mail: ]{m.kaufman@csuohio.edu}
\affiliation{$^1$Laboratoire de Physique Th\'eorique et Mod\'elisation \\
Universit\'e de Cergy-Pontoise, CNRS, UMR 8089 \\ 2, avenue Adolphe Chauvin,\\
95302 Cergy-Pontoise Cedex, France}
\affiliation{$^2$Permanent Address: Department of Physics, Cleveland State University, Cleveland, OH 44115, USA}


\date{\today}
\begin{abstract}
We study in this paper the phase transition in a mobile Potts model by the use of Monte Carlo simulation. The mobile Potts model is related to a diluted Potts model which is also studied here by a mean-field approximation. We consider a lattice where each site is either vacant or occupied by a $q$-state Potts spin. The Potts spin can move from one site to a nearby vacant site.  In order to study the surface sublimation, we consider a system of Potts spins contained in a recipient with a concentration $c$ defined as the ratio of the number of Potts spins $N_s$ to the total number of lattice sites $N_L=N_x\times N_y\times N_z$. Taking into account the attractive interaction between the nearest-neighboring Potts spins, we study the phase transition as functions of various physical parameters such as the temperature, the shape of the recipient and the spin concentration. We show that as the temperature increases, surface spins are detached from the solid phase to form a gas in the empty space.  Surface order parameters indicate different behaviors depending on the distance to the surface. At high temperatures, if the concentration is high enough, the interior spins undergo a first-order phase transition to an orientationally disordered phase. The mean-field results are shown as functions of temperature, pressure and chemical potential, which confirm in particular the first-order character of the transition.
\begin{description}
\vspace{0.5cm}
\item[PACS numbers: 05.50.+q ; 05.70.Fh ; 64.60.De  ]
\end{description}
\end{abstract}

\pacs{Valid PACS appear here}
\maketitle
\section{Introduction}
\label{sect:introduction}
Phase transition is a fascinating subject that has attracted an enormous number of investigations in various areas during the last fifty years.  Much of progress has been achieved in the seventies in the understanding of mechanisms which characterize a phase transition: the renormalization group shows that the nature of a phase transition depends on a few parameters such as the space dimension, the symmetry of the order parameter and the nature of the interaction between particles \cite{Wilson,Amit,Zinn}.

There has been recently a growing interest in using spin systems to describe properties of dimers and liquid crystals \cite{diep}. Spin systems are used in statistical physics to describe various systems where a mapping to a spin language is possible. In two dimensions (2D) Ising-Potts models were studied extensively \cite {Wu1982,Nienhuis,Kaufman}. Interesting results such as hybrid transitions on defect lines were predicted with renormalizaton group and confirmed with Monte Carlo (MC) simulations \cite {DiepKaufman}. Unfortunately, as for other systems of interacting particles, exact solutions can be obtained only for systems up to 2D with short-range interactions \cite{Baxter,Diep-Giacomini}.
We will focus in this paper on the $q$-state Potts model in three dimensions (3D) where Bazavova {\it et al.} have recently shown precise results for various values of $q$ for the localized Potts models \cite{Bazavova}.  High-temperature series expansions for random Potts models have been studied by Hellmund and co-workers \cite{Hellmund}. Other investigations have been carried out on critical properties in the 3D site-diluted Potts model \cite{Murtazaev} and Potts spin glasses \cite{Berker,Banavar,Gross,BerkerNishimori,Georgii}.

We are interested here in the problem of moving particles such as atoms or molecules in a crystal. To simplify, we consider the case of
mobile $q$-state Potts spins moving from one lattice site to a nearby one.
The $q$ states  express the number of internal degrees of freedom of each particle, such as molecular local orientations.
We simulate the mobile 6-state Potts model on a cubic lattice.  It is known that the pure Potts model in three dimensions undergoes a continuous transition for $q$ = 2 and a first-order transition for $q$=3, 4, ...
We use here MC simulation and a theoretical analysis to elucidate properties of such a system.  The mobility depends on the temperature.  At low temperatures, all spins gather in a solid, compact phase. As the temperature increases, spins at the surface are detached from the solid to go to the empty space forming a gaseous phase. We show that the phase transformation goes through several steps and depends on the concentration of the Potts spins in the crystal.

The mobile Potts model presented here is expected to be equivalent to the dilute Potts model in as far as the bulk thermodynamics is concerned. The kinematics at the interface between the solid and the gas phases may be affected by the constraint that atoms move only to empty neighboring cells in the mobile model as opposed to the case where atoms move to any other vacant cell in the diluted model.

At a sufficiently high concentration, spins are not entirely evaporated and the remaining solid core undergoes a transition to the orientationally disordered phase.
We anticipate here that there is only one phase transition in the model, a first-order transition from higher-density  (solid)  phase with non-zero Potts order parameter to a lower-density phase with vanishing Potts order parameter. The sublimation observed below is analogous to surface melting, which in the melting of a solid can begin well below the bulk melting temperature.
Details are shown and discussed in terms of surface sublimation and melting.
We note in passing that direct studies of melting using continuous atomic motions are efficient for bulk melting \cite{Gomez1,Gomez2,Bocchetti1} but they have often many difficulties to provide  clear results for complicated situations such as surface melting (see references cited in Ref. \onlinecite{Bocchetti2}). Using discrete spin displacements as in the present model we show that bulk melting and surface sublimation can be clearly observed. We believe that these results bear essential features of real systems.

Section \ref{sectionModel} is devoted to the description of our mobile Potts model.
The mobile Potts model is related to a diluted Potts model. The latter model is analyzed within the mean-field approximation for bulk properties in section \ref{sectionMF} below. The two models are not identical. While in the mobile Pots model a spin can move to a void location nearby, in the diluted model there is no constraint on the proximity of the locations of the spin and the vacancy. The thermodynamics of two models may be identical in the long run, even though the kinematics may be different.
Section \ref{sectionMC} is devoted to the presentation of MC simulation results. Concluding remarks are given in section \ref{sectionConcl}.

\section{Mobile Potts Model}\label{sectionModel}

We consider a lattice of $N_L$ sites. A site $i$ can be vacant or occupied at most by a Potts spin $\sigma_i$ of $q$ states: $\sigma_i=1,2,...,q$. Potts spins can move from one site to a neighboring vacant site under effects of mutual spin-spin interaction and/or of temperature $T$. In order to allow for spin mobility, the number $N_s$ of Potts spins should be smaller than $N_L$. Let us define the spin concentration $c$ by $c=N_s/N_L$. The Hamiltonian is given by the Potts model:
\begin{equation}\label{Hamil}
{\cal H}=-J\sum_{i,j}\delta (\sigma_i,\sigma_j)
\end{equation}
where $J$ is the interaction constant between nearest neighbors (NN) and the sum is taken over NN spin pairs. To this simple Hamiltonian, we can add a chemical potential term when we deal with the system in the grand-canonical description \cite{DiepSP} and an interaction term between neighboring vacancies (see below).

\begin{figure}[tb]
\begin{centering}
\includegraphics[width=0.25\textwidth]{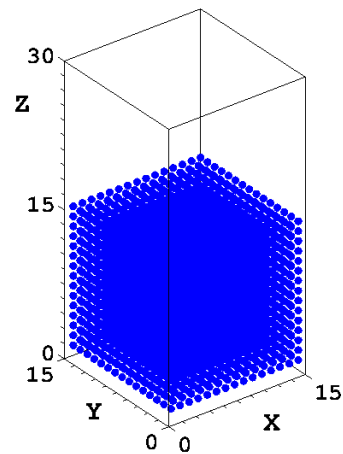}
\caption{(Color online) Ground state of the system with $N_x=N_y<N_z$.}
\label{figGS}
\end{centering}
\end{figure}

The ground state (GS) of the system described by Eq. (\ref{Hamil}) is the one with the minimum of interaction energy: each spin maximizes the number of NN of the same values. As a consequence, all spins  have the same value and form a compact solid. If the lattice is a recipient of dimension $N_x\times N_y\times N_z$,  then the GS is a solid with a minimum of surface spins (surface spins have higher energies than interior spins due to a smaller number of NN). In a recipient with  $N_x=N_y<N_z$ with periodic boundary conditions in the $xy$ plane and close limits on the $z$ direction for example, the free surface is the $xy$ surface.  We show in Fig. \ref{figGS} such an example.

When $T$ is increased, surface spins are detached from the solid to go to the empty space.  At high $T$, the solid becomes a gas.  The path to go to the final gaseous phase will be shown in this paper. We start with the bulk case and examine the surface behavior in what follows.

\section{Mean-Field Theory}\label{sectionMF}

In this section, we present the mean-field theory for the mobile Potts model. It is more convenient to work in the grand-canonical description. The results do not depend on the approaches for large systems \cite{DiepSP}. To that end we consider a vacancy as a spin with value zero. The model becomes $(q+1)$-state model. In addition, we add a chemical term in the Hamiltonian and rewrite it in a more general manner in the following.

\subsection{Hamiltonian}

We divide the space into $M$  cells, of equal volume $v$,  centered each on a site of a cubic lattice. Any cell is either vacant or occupied by a single particle characterized by a $q$-value spin.  Neighboring particles that have the same spin value get a lower interaction energy $-J$ than if they have different spin values.  Zero energy is assigned to neighboring cells that have at least a vacancy.  We assign an energy $-K$ to neighboring cells that are occupied irrespective of their spin values.
In the grand canonical ensemble we allow for fluctuating number of particles and include in the Hamiltonian a single site (cell) term proportional to the chemical potential $H$ if there is a particle at the cell.  This model can be described by assigning at each site a $(q + 1)$-Potts spin $\sigma= 0, 1,…,q$. The zero value corresponds to vacancy while the values 1, 2,..., $q$ correspond to a particle having a spin.
The Hamiltonian is
\begin{eqnarray}
-\frac{{\cal H}}{k_B T}&=&J\sum_{i,j}\delta(\sigma_i,\sigma_j) [1-\delta (\sigma_i,0)][1-\delta (\sigma_j,0)]\nonumber\\
&&+K\sum_{i,j}[ 1-\delta (\sigma_i,0)][1-\delta (\sigma_j,0)]\nonumber\\
&&+H\sum_{i}[ 1-\delta (\sigma_i,0)]\label{eq1}
\end{eqnarray}
This corresponds to the grand canonical ensemble: fixed temperature $T$, chemical potential $H = \mu/k_BT$ and volume $V$.

\subsection{Mean-Field Theory}
The mean-field theory of the diluted Potts model \cite{Kaufman-Kardar} is exact for the equivalent-neighbor lattice.  The thermodynamic potential divided by $M$ is proportional to the pressure:
\begin{equation}\label{eq2}
-pv=\frac{\Omega}{M} = -k_B T \frac{\ln Z}{M}=k_B T \min(\Psi)
\end{equation}
where
\begin{eqnarray}
\Psi &=&\frac{J}{2 } (m_1^2+m_2^2…+m_q^2+...)\nonumber\\
&&-\ln (1+e^{Jm_1+H}+e^{Jm_2+H}+...+e^{Jm_q+H})\label{eq3}
\end{eqnarray}
Note that the above equations is for $K = 0$.
The optimization equations are
\begin{equation}\label{eq4}
m_a= \frac{ e^{Jm_a+H}}{1+e^{Jm_1+H}+e^{Jm_2+H}+...+e^{Jm_q+H} }
\end{equation}
for $a = 1, ..., q.$							
The $m_a$ gives the average number of particles of spin $a$ normalized by the total number of sites (cells) $M$.
The number of particles normalized by $M$ is
\begin{equation}\label{eq5}
n= \sum_{a=1}^q m_a = \sum_{a=1}^q \frac{e^{Jm_a+H}}{1+e^{Jm_1+H}+e^{Jm_2+H}+...+e^{Jm_q+H}}							 \end{equation}
Assuming the ordering of the Potts spin to occur in state 1, we parameterize the $m'$s as follows
\begin{equation}\label{eq6}
m_1=  \frac{n}{q}+(q-1)m\ \ ;\ \ m_2= m_3=...= m_q=\frac{n}{q}-m						 \end{equation}
The optimization equations (\ref{eq4})-(\ref{eq5}) are now
\begin{eqnarray}
\frac{qm}{n}&=&\frac{e^{Jqm}-1}{e^{Jqm}+q-1}	\label{eq7}\\
n&=&\frac{e^{J(n/q-m)+H} (e^{Jqm}+q-1)}{e^{J(n/q-m)+H} (e^{Jqm}+q-1)+1}\label{eq8}
\end{eqnarray}
In the following we denote $mq/n = X$.  The energy $U$ scaled by $M$, number of cells, is
\begin{equation}\label{eq9}
U=-\frac{1}{2}\sum_{a=1}^q m_a^2=-\frac{n^2}{2q}-\frac{ (q-1) n^2 X^2}{2q}	
\end{equation}							
The specific heat at fixed number of particles $n$ is
\begin{equation}\label{eq10}
C_v=\frac{dU}{dT}=-\frac{q-1}{2q} n^2 \frac{dX^2}{dT}
\end{equation}
The second optimization equation, Eq. (\ref{eq8}), provides a formula for the chemical potential, since $H = \mu/T$,
\begin{equation}\label{eq11}
\mu=T \ln\frac{n}{1-n}+T \ln \frac{1-X}{q}-\frac{n(1-X)}{q}
\end{equation}
The pressure $p$ is obtained from Eq. (\ref{eq2})
\begin{equation}\label{eq12}
pv=-\frac{n^2}{2q} [1+(q-1) X^2 ]-T \ln (1-n)
\end{equation}
Note in the disordered (gas) phase $X = 0$ and $n << 1$.  The equation of state reduces to the ideal gas equation $pv = nT$.
The entropy $S$ normalized by $M$ is obtained from the thermodynamic Euler equation
\begin{eqnarray}
S&=&\frac{u+pv-\mu n}{T}= -n\ln(n)-(1-n) \ln(1-n)\nonumber\\
&&-n\ln \frac{1-X}{q}-\frac{n^2}{qT}[(q-1) X^2+X]\label{eq13}
\end{eqnarray}

\begin{figure}[ht!]
\centering
\includegraphics[width=8cm,angle=0]{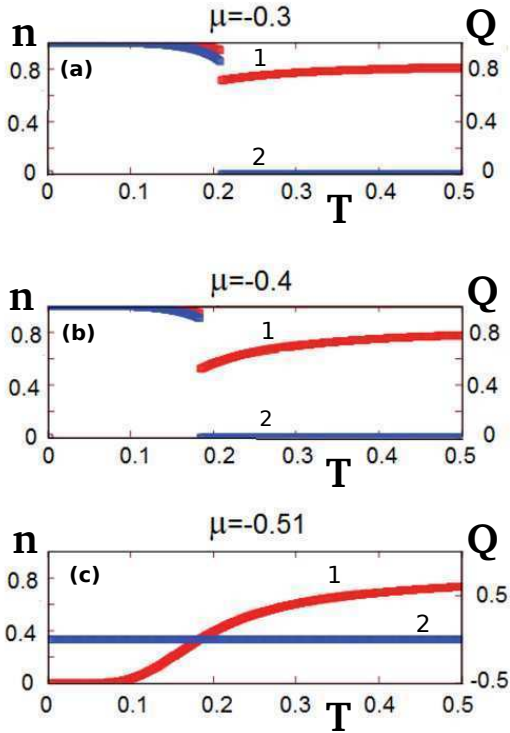}
\caption{(Color online) Average number of particles per site $n$ (curve 1, red, left scale) and order parameter $Q$ (curve 2, blue, right scale) versus temperature $T$ with (a) $\mu=-0.3$, (b) $\mu=-0.4$, (c) $\mu=-0.51$. See text for comments.\label{figmk1}}
\end{figure}

The model exhibits a first-order phase transition tied to the Potts $q$-state transition. In Fig. \ref{figmk1} we show  $n$ by curve 1 (red) and the order parameter $Q = qm$ [see Eq. (7)] by curve 2 (blue) as functions of $T$ for fixed chemical potential $\mu = -0.4$ (Fig. \ref{figmk1}b).
Increasing the chemical potential reduces the discontinuity in $n$ as seen  for $\mu = -0.3$ (Fig. \ref{figmk1}a). While, decreasing the chemical potential below $\mu=-0.5$ destroys order at all temperatures as seen for $\mu=-0.51$ (Fig. \ref{figmk1}c).

This is understood by comparing the energy of any pair $(i,j)$ in the ordered (solid) phase $E_{i,j} = J +2H$ to the energy in the disordered (gaseous) phase $E_{i,j} = 0$. The two energies cross when $H = -0.5J$, or when $\mu = - 0.5$.

\begin{figure}[ht!]
\centering
\includegraphics[width=6cm,angle=0]{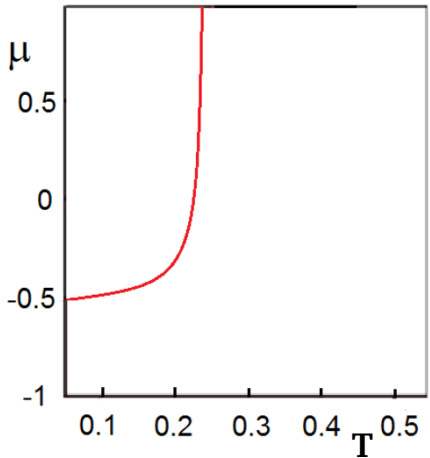}
\caption{(Color online) Phase diagram in the plane ($T,\mu$). The solid line is a first-order transition line.\label{figmk4}}
\end{figure}

The phase diagram in the ($T,\mu$) plane shown in Fig. \ref{figmk4} includes an ordered (solid) phase (low $T$ and high $\mu$) and a disordered (gas) phase. The two phases are separated by a line of first-order transitions.

In the limit of large chemical potential the number of vacancies becomes negligible and thus the model reduces to the $q$-state Potts model. As a result the transition line approaches $T = 0.25$.

\begin{figure}[ht!]
\centering
\includegraphics[width=6cm,angle=0]{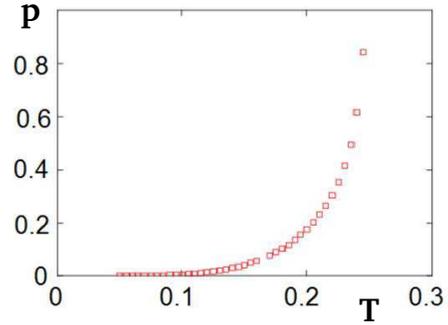}
\caption{(Color online) Phase diagram in the plane ($T,p$). \label{figmk5}}
\end{figure}

The phase diagram in the temperature-pressure plane is shown in Fig. \ref{figmk5}.

\begin{figure}[ht!]
\centering
\includegraphics[width=6cm,angle=0]{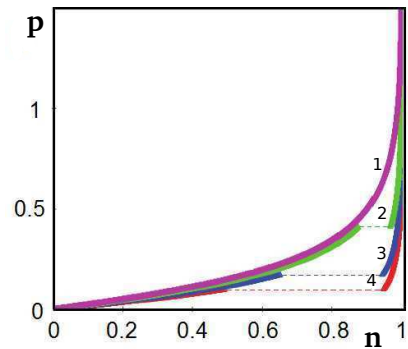}
\caption{(Color online) Isotherms ($p,n$) are shown by thick curves for $T=0.25$ (curve 1, violet), 0.23 (curve 2, green), 0.2 (curve 3, blue) and 0.18 (curve 4, red). Thin broken lines indicate discontinuities of $n$.\label{figmk6}}
\end{figure}

Isotherms pressure $p$ vs $n$ are shown in Fig. \ref{figmk6}. For $T = 0.25$ there is no phase transition while for $T= 0.23$, 0.2, 0.18 the first order-transitions line is crossed. As a result, we see the gap in the density $n$.

\begin{figure}[ht!]
\centering
\includegraphics[width=7cm,angle=0]{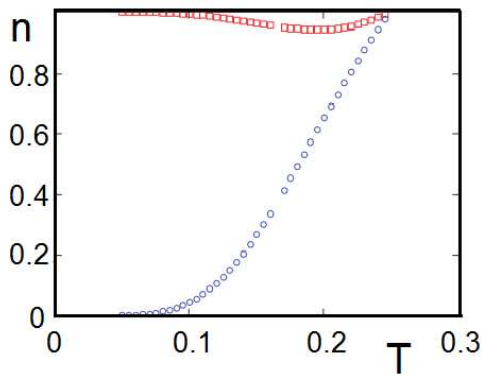}
\caption{(Color online) Phase diagram in the plane ($T,n$).  See text for comments.\label{figmk7}}
\end{figure}

The higher $n$ branch corresponds to the ordered solid.  Note that the solid exists only for $n$ large enough ($n > 0.95$).
In other words the presence of 5\% vacancies destroys the solid.  This is summarized in the phase diagram in the ($T,n$) plane shown in Fig. \ref{figmk7}. The two lines represent the densities of the solid (red squares) and of the gas (blue circles). The two branches coalesce at a temperature of 0.25. Note that this is not a critical point but the end of the thermodynamic space as it occurs in the limit of infinite chemical potential (i.e. no vacancies).

\begin{figure}[ht!]
\centering
\includegraphics[width=8cm,angle=0]{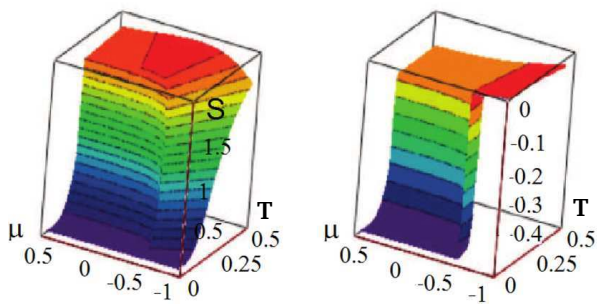}
\caption{(Color online) Entropy (left) and energy (right) as functions of temperature and chemical potential.\label{figmk8}}
\end{figure}

Entropy and energy versus temperature and chemical potential are shown in Fig. \ref{figmk8}.

\begin{figure}[ht!]
\centering
\includegraphics[width=7cm,angle=0]{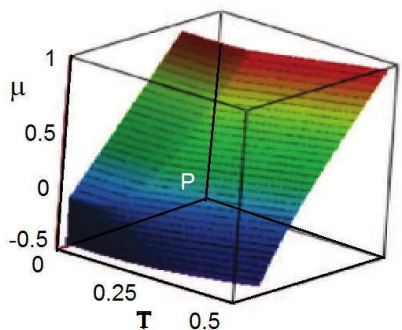}
\caption{(Color online) Surface in the space ``temperature, pressure and chemical potential".\label{figmk10}}
\end{figure}

The fundamental equation, chemical potential as a function of temperature and pressure, is concave as required by the second law of thermodynamics (thermodynamic stability). It is a continuous function and the first-order transitions manifest as discontinuities in slope of the chemical potential when graphed against temperature and pressure (Fig. \ref{figmk10}).

\section{Monte Carlo results}\label{sectionMC}
In this section, we present our results from MC simulations. The method can be briefly described as follows. At a given $T$, we take a spin and calculate its interaction energy with its NN. We then move it to one of nearby vacant sites chosen at random, change its state chosen at random among $q$ states. We calculate its ``new energy". If this is lower than its old energy, then the new spin state and new position are accepted. Otherwise, we use the Metropolis criterion \cite{Binder} to accept or reject its new situation.  We repeat this update procedure for all spins: such a system sweeping is called one MC step (MCS).

In our simulations, we used $10^5$ MCS/spin to equilibrate the system before averaging physical quantities over the following $10^6$ MCS/spin.  We have verified that longer MC run times do not change the results. We used various system sizes and shapes to examine finite-size and shape effects on the results.

\subsection{Transition}

We study here the melting behavior of a solid contained in a recipient described in section \ref{sectionModel}. The recipient has the dimension $N_x=N_y<N_z$ and is filled with molecules (Potts spins) in the lower part of the recipient. The number of filled layers is smaller than $N_z$ (Fig. \ref{figGS}). Molecules under thermal effect can be evaporated from the upper surface to the empty space.
To study the behavior of such a system, we choose to heat the system from low to high $T$. Cooling the system from a random initial configuration, namely molecules in a gas state with positions distributed over all space, will result in a compact solid phase at low temperature but the surface of this solid is not so flat so that the system energy is about $5\%$ higher than the GS energy shown in Fig. \ref{figGS}. However, the system behavior at higher $T$ as well as the phase transition  are the same as obtained by heating. We will show this later.

Let us show now results for a lattice of $15 \times  15 \times 30$ sites where only the fifteen first layers in the $z$ direction
are filled ($c=50\%$) in the GS configuration shown in Fig. \ref{figGS}. As said above, this configuration corresponds to the one with a minimal free surface when the system is in the solid state.

Our simulation in real time shows that when $T$ increases atoms on the surface are progressively evaporated.  The solid core of the system remains in a Potts spin order, though its volume is little by little reduced with increasing $T$. At a high enough value of $T$, say $T_c$,  the Potts orientational order of the solid core is broken. However, the spins still stay in the solid state up to a very high $T$ when the whole system melts to a gas (or liquid) phase.  We will show later evidence of such a change of the system at several $T$ with snapshots and corresponding distributions of the NN number.

\begin{figure}
\centering
\includegraphics[width=0.35\textwidth]{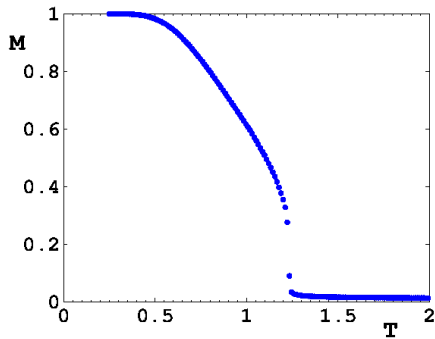}
\caption{(Color online) Magnetization $M$ versus  temperature $T$ (in unit of $J/k_B$) for a lattice of $15 \times  15 \times 30$ sites with a spin concentration $c=50\%$.  The system is completely ordered at $T=0$, then surface spins are little by little evaporated with increasing $T$.  The phase transition of spin orientations of the solid core occurs at $T_c\simeq 1.234$.}
\label{fig:M_15x15x30_50pc}
\end{figure}

 The magnetization $M$ versus $T$ is displayed in Fig.  \ref{fig:M_15x15x30_50pc} where $M$ indicates a perfect order at low $T$. When $T$ is increased $M$ decreases linearly with $T$: a careful examination of the system dynamics reveals that this regime corresponds to the evaporation of surface spins. This regime ends with a discontinuity of $M$ at a transition temperature $T_c\simeq 1.234$.

\begin{figure}[tb]
\centering
\includegraphics[width=0.35\textwidth]{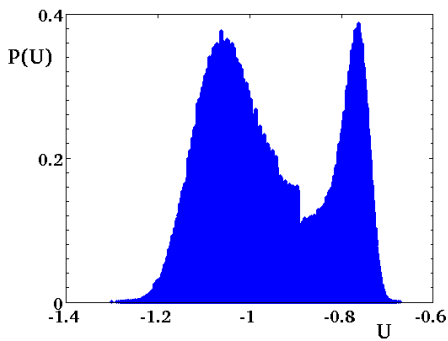}
\caption{(Color online) Energy histogram $P(U)$ recorded at $T_c=1.234$ for a lattice $15 \times  15 \times 30$ with $c=50\%$.
The presence of the two peaks indicates that the transition is of first order.}
\label{fig:histo}
\end{figure}

The discontinuity at $T_c$ indicates a first-order phase transition. We have verified this by recording the energy histogram $P(U)$ at the transition temperature $T_c=1.234$. The double-peak structure shown in Fig. \ref{fig:histo} confirms the first-order character of the transition.
Note that disordered evaporated atoms, namely atoms outside the system solid core, do not participate in the transition.

\begin{figure}[tb]
\centering
\includegraphics[width=0.35\textwidth]{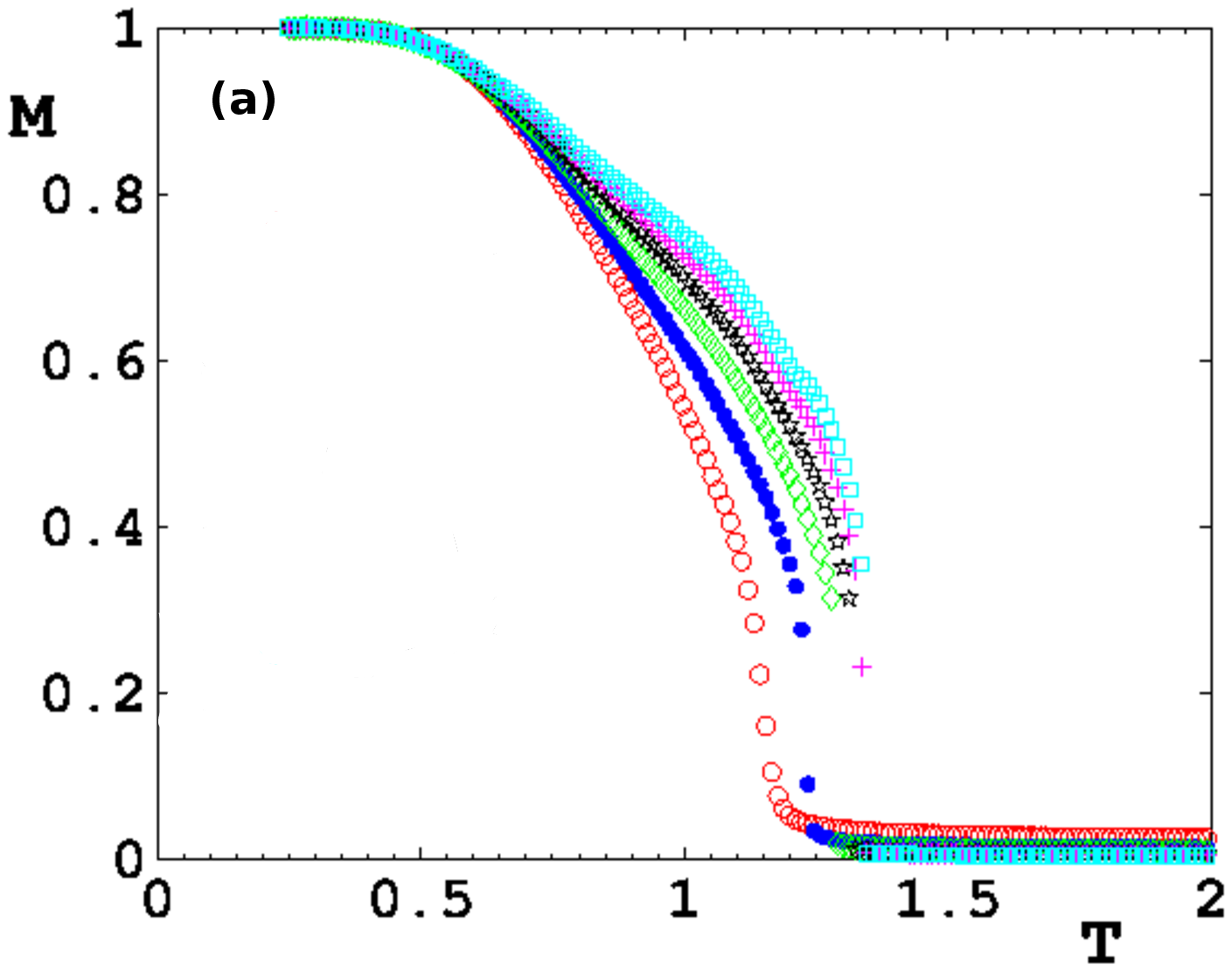}
\includegraphics[width=0.35\textwidth]{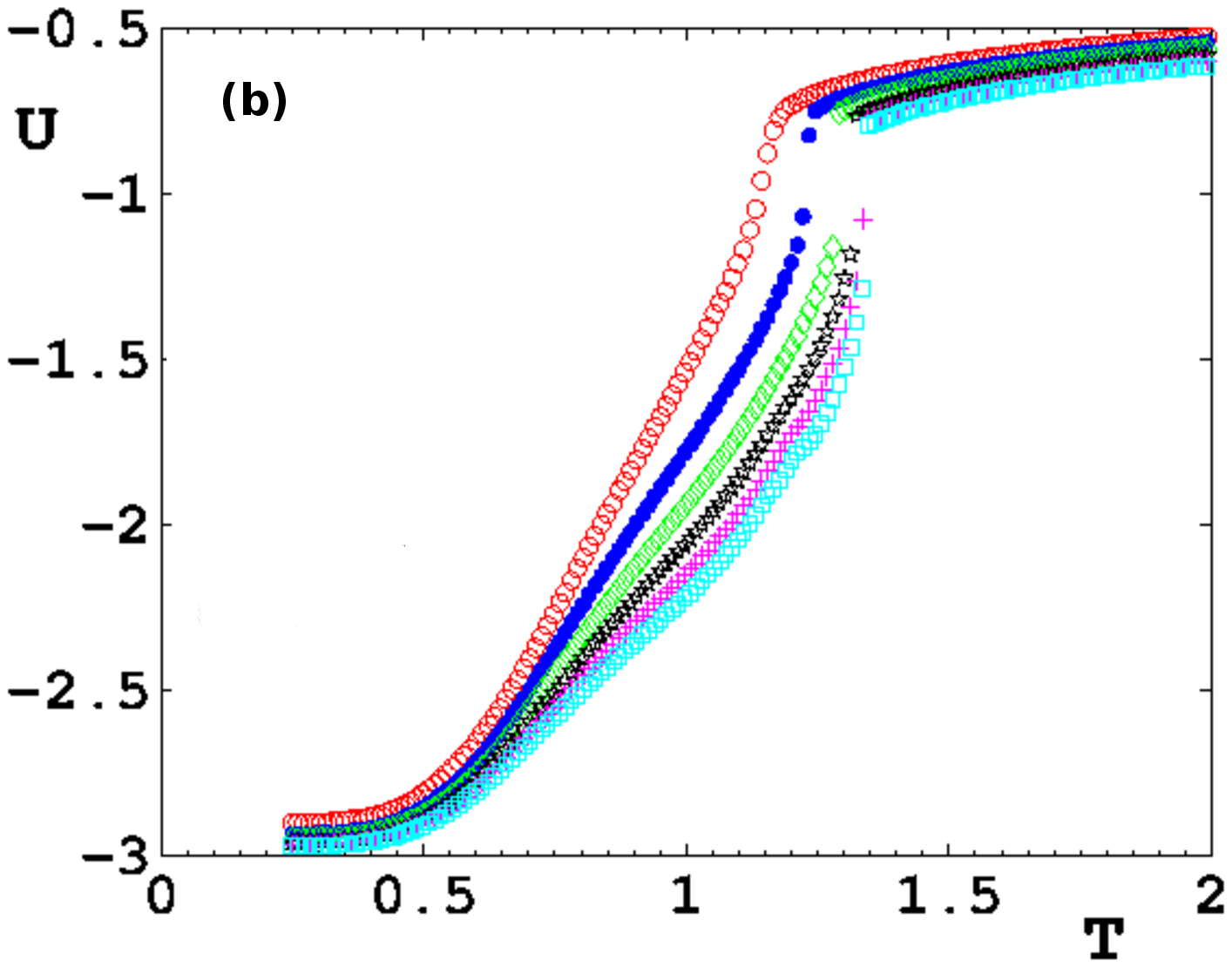}
\caption{(Color online)  Comparison of the evolution of (a) the magnetization
and (b) the energy versus the temperature of a  half-filled lattice  $N_x \times  N_y \times N_z$ for several sizes $N_x=N_y=N_z/2=20$ (red void  circles), 30 (blue filled circles), 40 (green void diamonds), 50 (black stars), 60 (magenta crosses) and 70 (sky-blue void squares).}
\label{fig:Size_effect_50pc}
\end{figure}

We have studied the finite-size effect on the transition at $c=50\%$. Since the shape of the recipient has a strong effect on the phase transition as seen below, we have kept the same recipient shape to investigate the finite-size effect: to compare results at the same concentration with those of the lattice $15 \times  15 \times 30$ sites, we have used lattices of $20 \times  20 \times 40$,  $25 \times  25 \times 50$,  $30 \times  30 \times 60$ and $35 \times  35 \times 70$ sites in which half of the recipient is filled with  spins, namely $c=50\%$.  We show in Fig. \ref{fig:Size_effect_50pc} the magnetization and the energy versus $T$ for several sizes.

\begin{figure}[tb]
\centering
\includegraphics[width=0.35\textwidth]{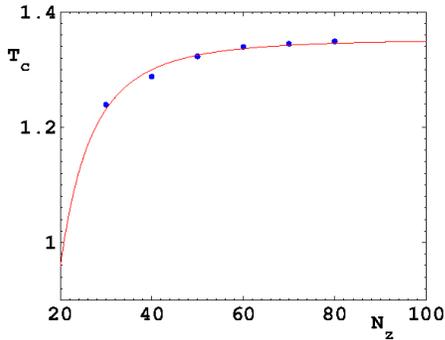}
\caption{(Color online) Transition temperature versus lattice size
$N_x \times  N_y \times N_z$ at $c=50\%$ where $N_x=N_y=N_z/2$, with $N_z=30$, 40, 50, 60, 70 and 80.}
\label{fig:Tc}
\end{figure}

Figure \ref{fig:Size_effect_50pc} shows that the transition looks like a second-order transition when the size of the
box is small. This is a well-known finite-size effect: when the linear size of a system is smaller than the correlation length at the transition, the system behaves as  a second-order transition.  We have to use therefore a finite-size scaling to ensure that the transition is of first order. To do this, let us show the transition temperatures for  systems at various sizes in Fig. \ref{fig:Tc}.
By fitting simulation results with the finite-scaling formula
\[T_c(L)=T_c(\infty)+\frac{A}{L^{\alpha}}\]
we find the following best nonlinear least mean square fit with the relative change of the last (8th) iteration    less than $-1.30954\times 10^{-10}$:

\[T_c(\infty)=1.35256 \pm 0.004089\ \  (0.3023\%),\]
\[\alpha=2.9,\ \quad A= -2335.24 \pm 170.9\ \   (7.316\%).\]
This is shown by the continued line in Fig. \ref{fig:Tc}.

Several remarks are in order:
 (i) the value of $\alpha$ indicates that, within statistical errors, $T_c(L)$ does scale with the system volume $L^3$ as it should for a first-order transition \cite{Barber,Bazavova},
 (ii) our value of $T_c(\infty)$ is in excellent agreement with that found for the localized model $T_c=1.35242\pm 0.00001$ obtained with the state-of-the-art multi-canonical method \cite{Bazavova} with periodic boundary conditions in three directions (note that in the original paper the authors have used a factor 2 in the Hamiltonian),
 (iii)  the fact that our system follows the same finite-size scaling as the localized model confirms that the transition observed in our mobile model is triggered by the orientational disordering of Potts spins in the remaining solid core at the transition temperature.

 Following the last argument, it is then obvious that if the quantity of matter remaining in the solid phase is so small  due to the evaporation, then there is no transition.  This should be seen if we lower the concentration.

\begin{figure}[tb]
\centering
\includegraphics[width=0.40\textwidth]{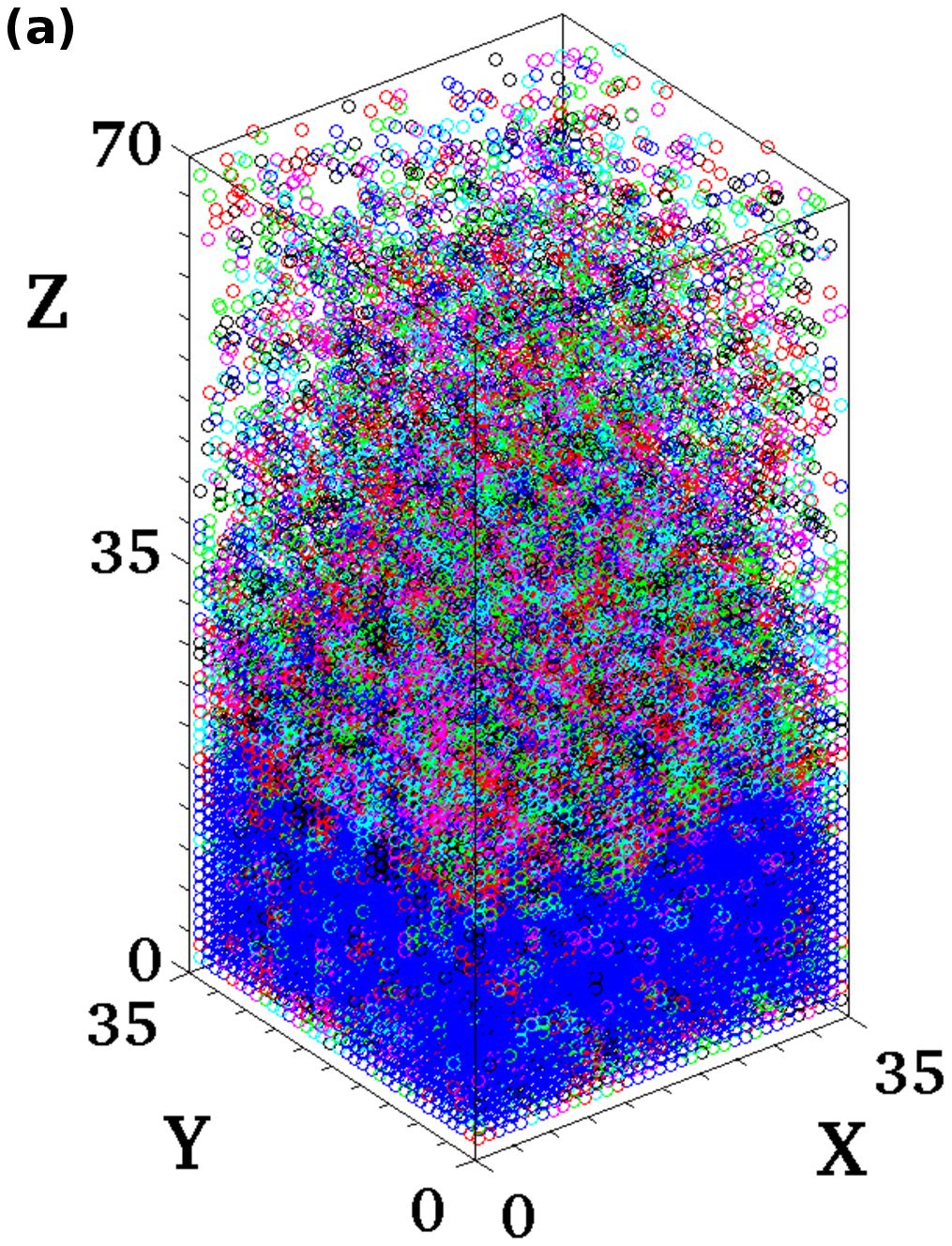}
\includegraphics[width=0.35\textwidth]{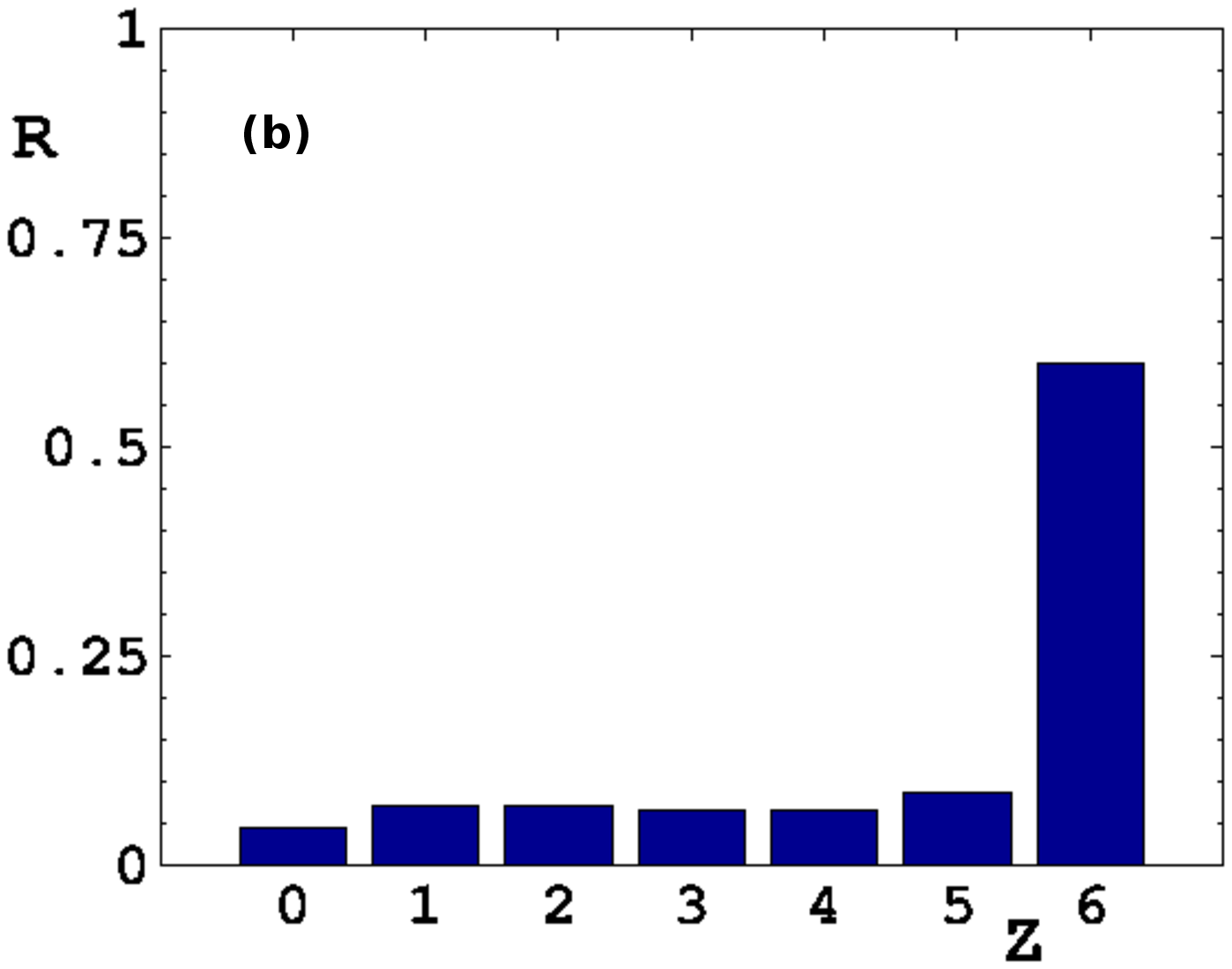}
\caption{(Color online)  Simulation for a half filled lattice size $35 \times 35 \times 70$ ($c=50\%$):
(a) Snapshot at $T=1.3128$ close to the transition,
 (b) $R$ vs $Z$, $R$ being the percentage of lattice sites having $Z$ nearest neighbors, at $T=1.3128$. Note that the solid phase is well indicated by the number of sites with 6 neighbors. }
\label{fig:Snap1.3}
\end{figure}

Before showing the effect of concentration, let us show a snapshot in the case of $c=50\%$ in Fig. \ref{fig:Snap1.3}.  The transition scenario discussed above is seen in these snapshots: the observed transition is that of Potts orientational order in the solid core.


Note that unlike the crystal melting where atoms suddenly quit their low-$T$ equilibrium positions to be in a liquid state, our model shows that the passage to the gaseous phase takes place progressively with slow evaporation, atom by atom, with increasing $T$.

Before showing the concentration effect let us compare results of heating and cooling.  As said above, cooling the system from an initial configuration where molecules are distributed at random over the whole space results in a compact solid phase at low $T$ shown in Fig. \ref{figGS}. However, this is realized only if we do a slow cooling: the final configuration at a temperature is used as initial configuration for a little bit lower temperature and so on. A rapid cooling will result in a solid with an irregular form having flat surfaces of various sizes.


\subsection{Effect of concentration}
\begin{figure}[tb]
\centering
\includegraphics[width=0.35\textwidth]{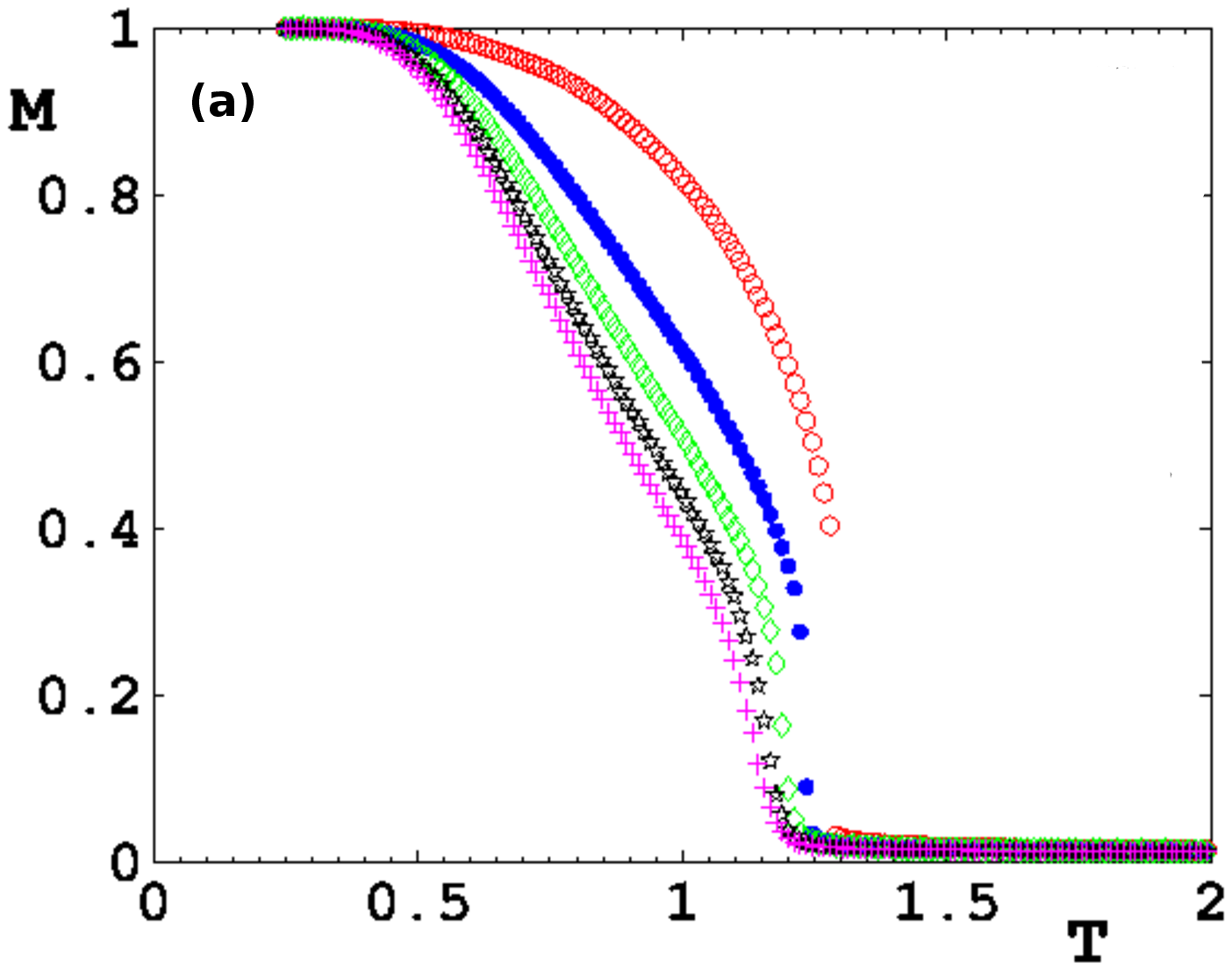}
\includegraphics[width=0.35\textwidth]{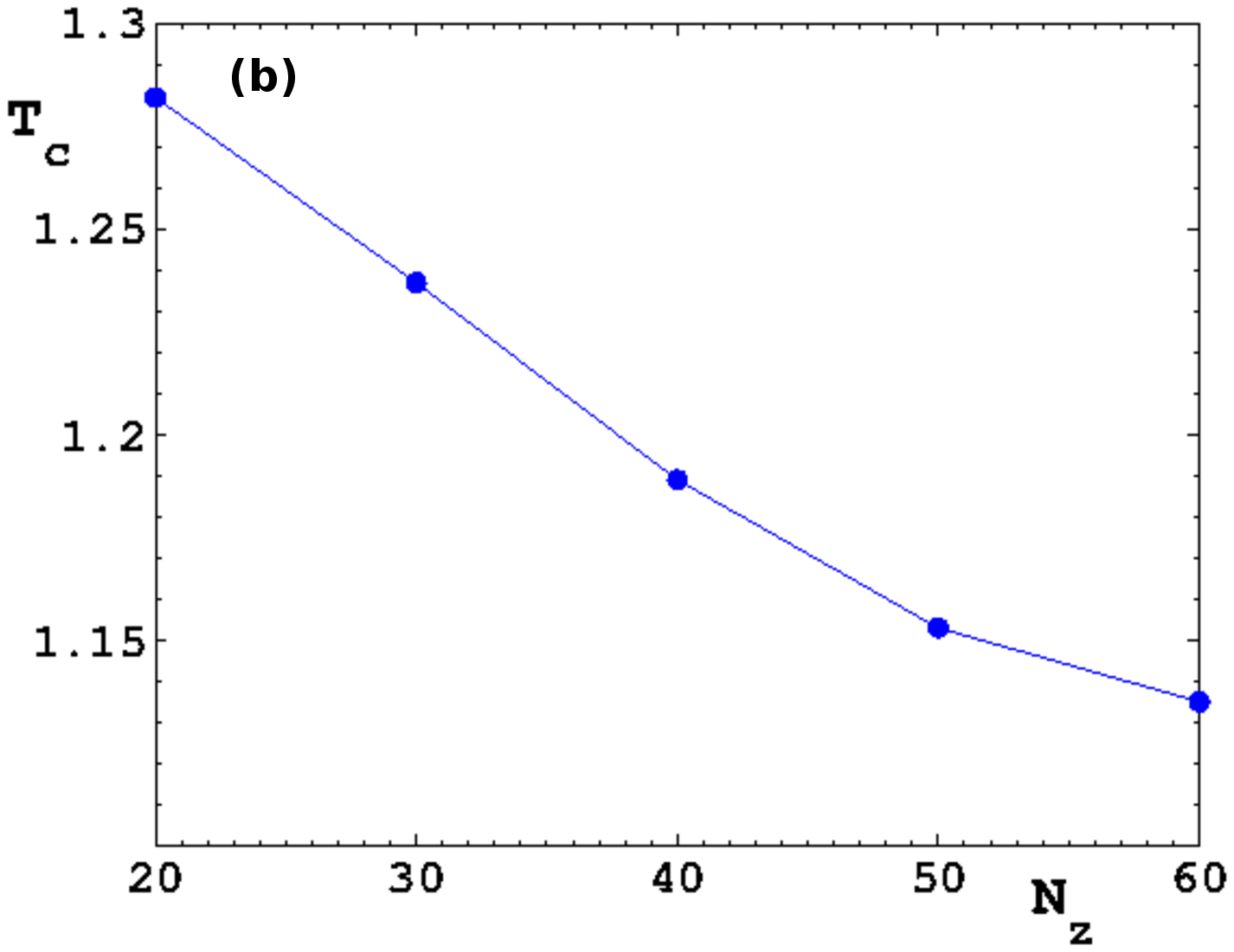}
\caption{(Color online) Effect of concentration: (a) Magnetization versus temperature for a lattice $15 \times  15 \times N_z$ where $N_z=20$ (red crosses), 30 (black stars), 40 (green diamonds), 50 (blue filled circles), 60 (red void circles). For each case, only the fifteen first layers are filled, corresponding to concentrations $15/N_z$, (b) Transition temperature versus the recipient height $N_z$.}
\label{fig:15x15xNz}
\end{figure}

Let us examine now results of simulations with smaller concentrations.
The absence of the phase transition is seen when we decrease the concentration down to $c=20\%$. As we can see in Fig. \ref{fig:15x15xNz}, at low temperatures, in all cases the system is in a condensed state. As $T$ increases, the magnetization decreases faster at lower concentrations.
All atoms are evaporated for small concentrations below the Potts transition temperature.
\\

Note, however, that for low concentrations, there is no transition but the magnetization disappears only when the  very small solid core disappears, namely at $T\simeq 1.1$.

\subsection{Surface sublimation}

\begin{figure}[h!]
\centering
\includegraphics[scale=0.38]{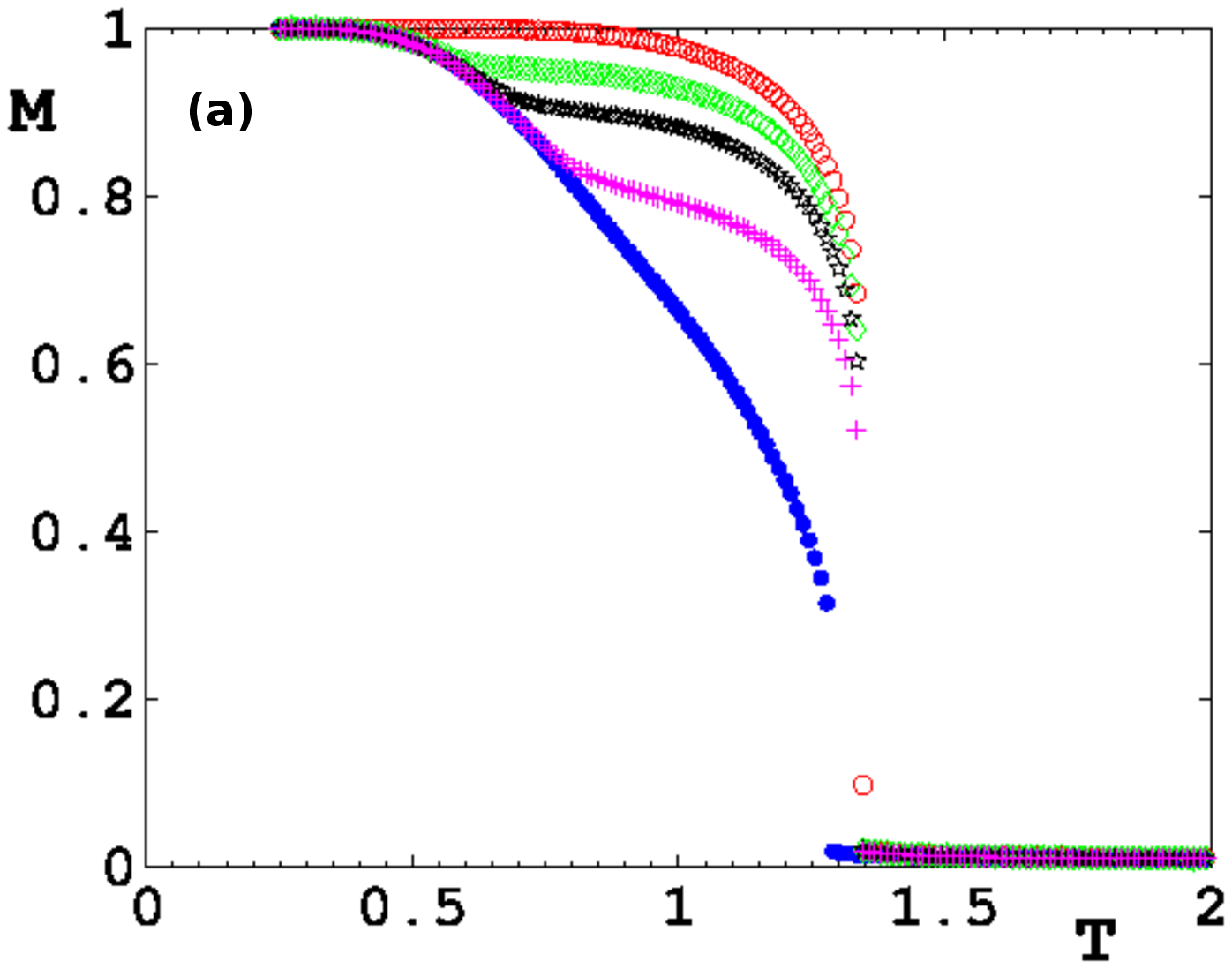}
\includegraphics[scale=0.38]{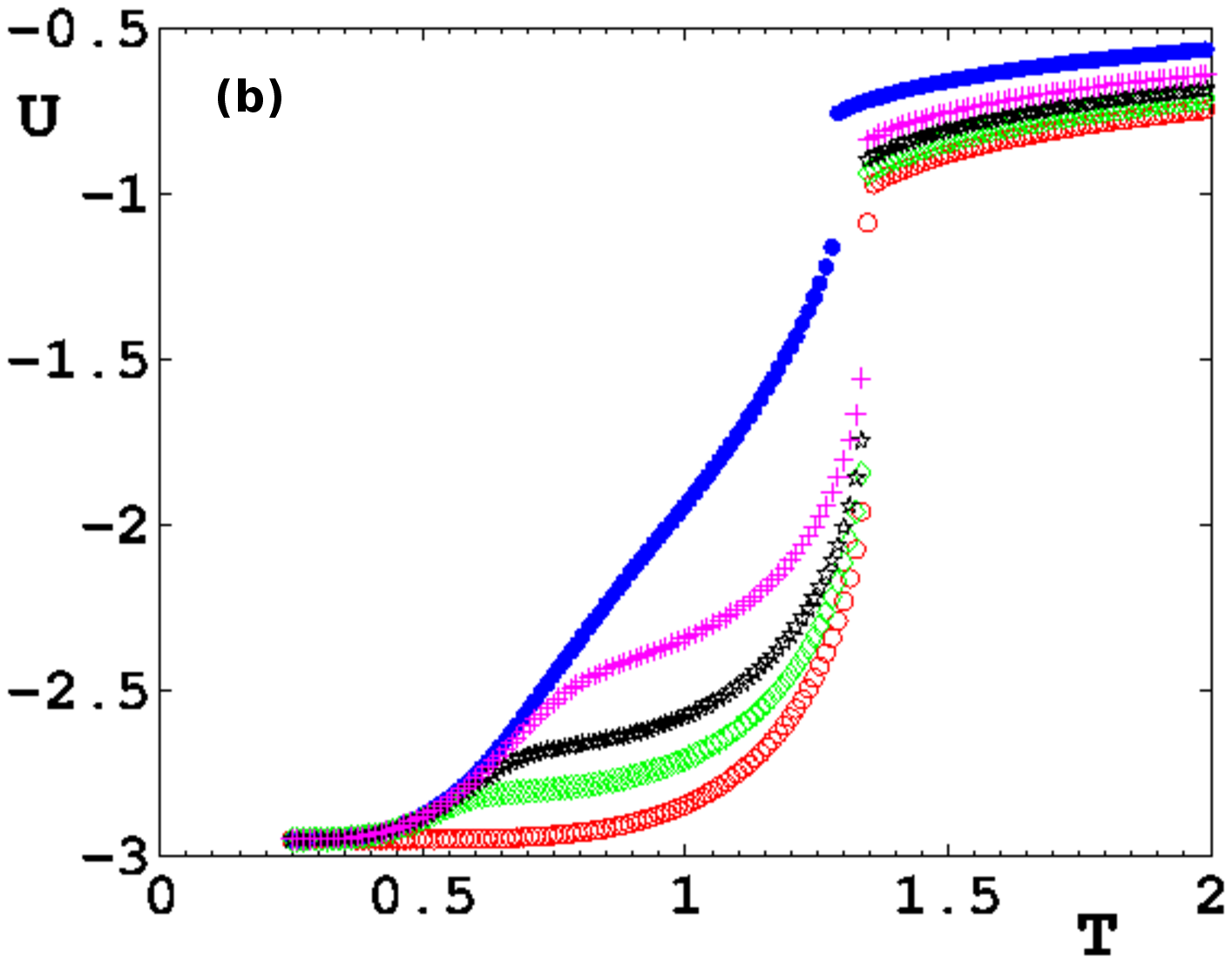}
\caption{(Color online) (a) Total magnetization $M$ and (b) energy per spin $U$ versus $T$. Red void circles indicate results of localized spins and blue filled circles indicate those of the completely mobile model.  Between these two limits, green void diamonds, black stars and magenta crosses correspond respectively to the cases where one, two and four surface layers are allowed to be mobile. }\label{figMtot}
\end{figure}

Let us show the results using the system size $20 \times 20 \times 40$ with $c=50\%$. To appreciate the surface sublimation, we show also the results of the localized model where spins stay each on its site. Figure \ref{figMtot} shows the total magnetization and the energy per spin.



\begin{figure}[h!]
\centering
\includegraphics[scale=0.38]{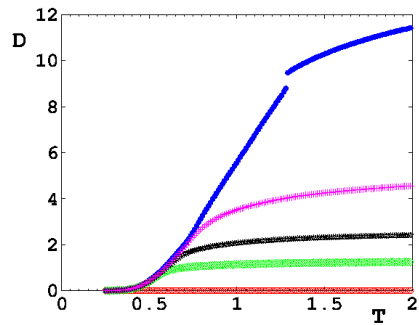}
\caption{(Color online) Diffusion coefficient $D$ versus $T$. Red void circles (lowest curve) indicate results of localized spins and blue filled circles (topmost curve) indicate those of the completely mobile model.  Between these two limits, from below green void diamonds, black stars and magenta crosses correspond respectively to the cases where one, two and four surface layers are allowed to be mobile.}\label{figCdiff}
\end{figure}

We show in Fig. \ref{figCdiff} the diffusion coefficient $D$ for the cases where one, two and four layers are allowed to be mobile. As seen, the evaporation is signaled by the change of curvature of $D$. Only when all layers are allowed to be mobile that the transition becomes really of first order with a discontinuity.

\begin{figure}[h!]
\centering
\includegraphics[scale=0.38]{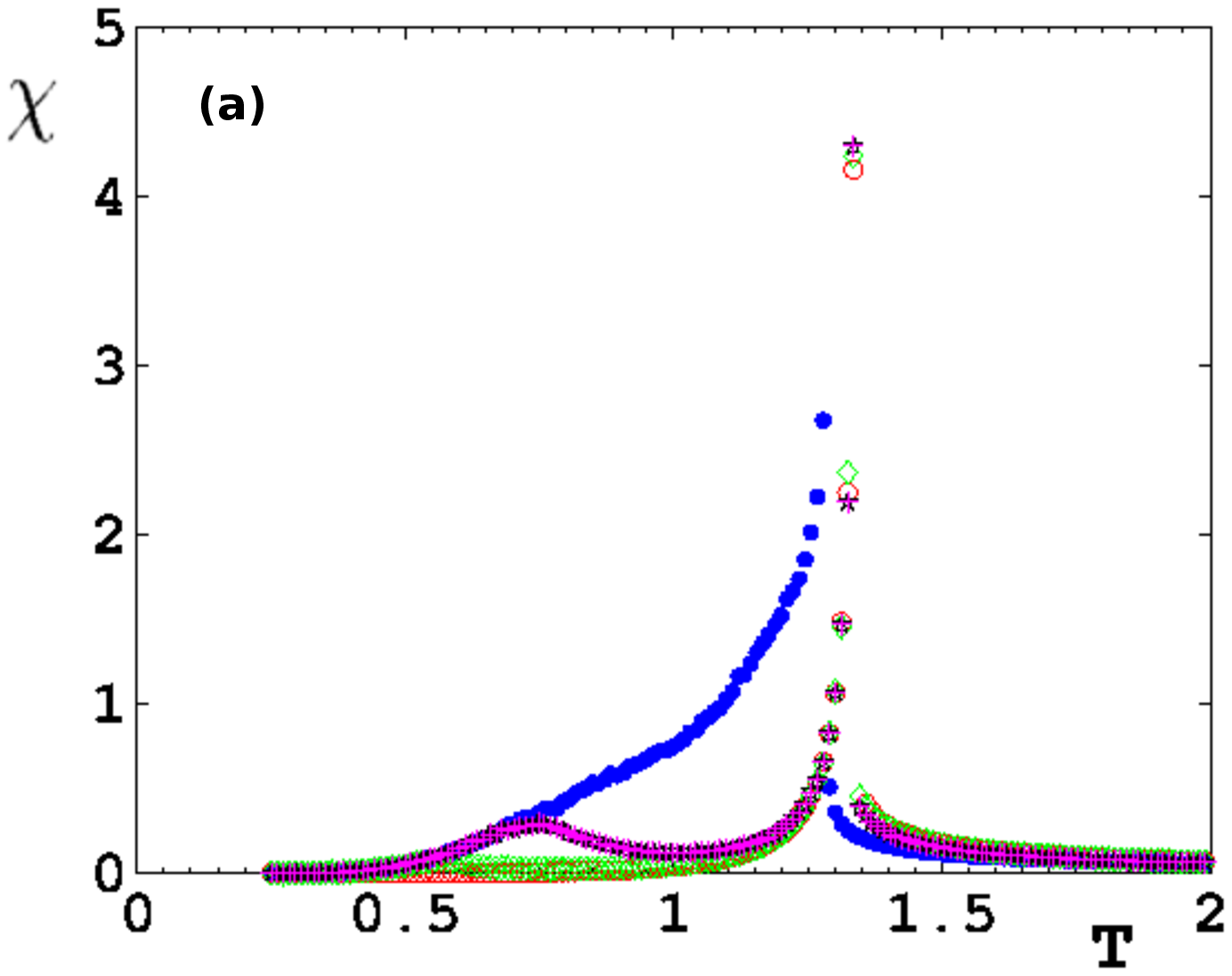}
\includegraphics[scale=0.38]{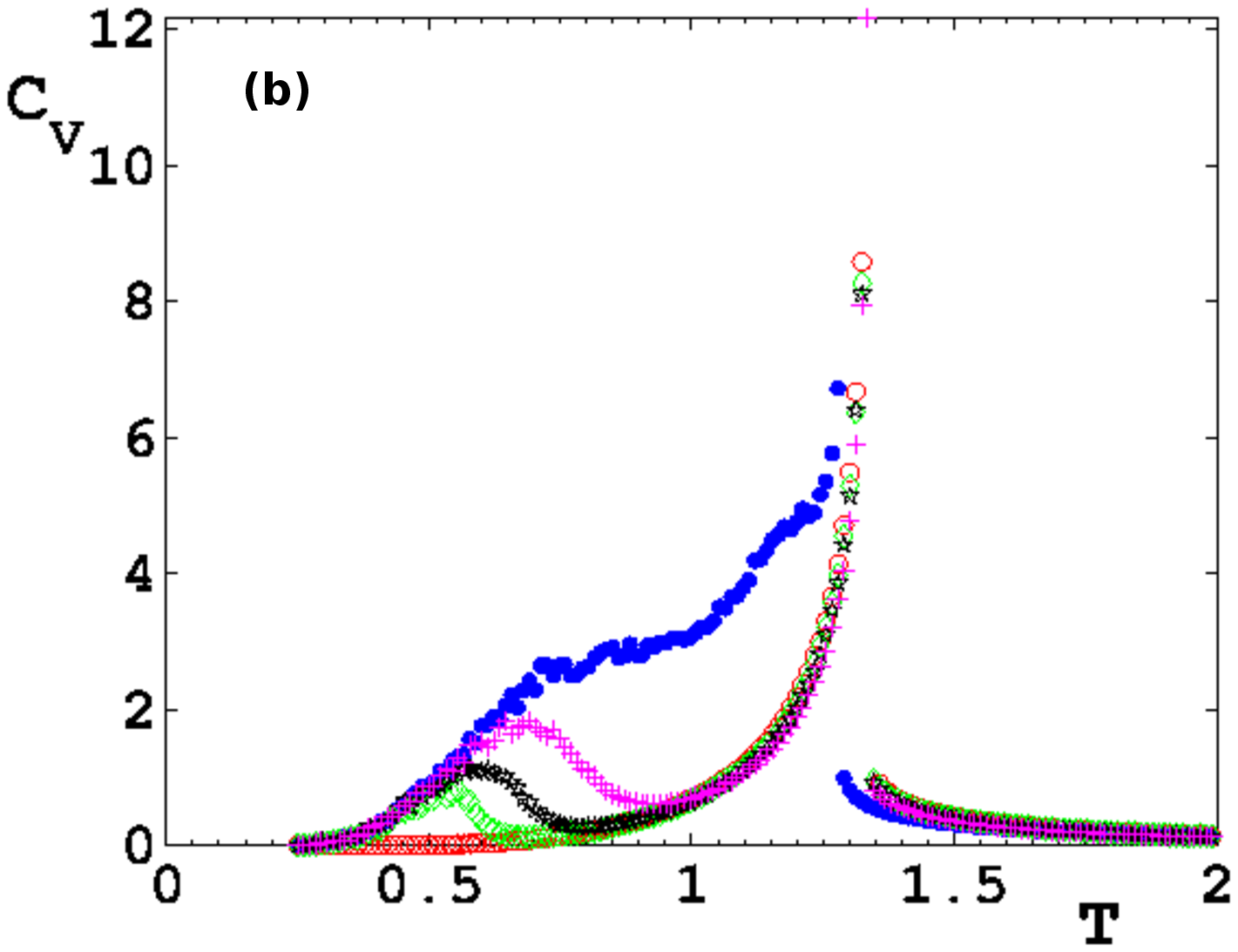}
\caption{(Color online) (a) Magnetic susceptibility $\chi$ and (b) heat capacity $C_V$ versus $T$. Red void circles indicate results of localized spins and blue filled circles indicate those of the completely mobile model.  Between these two limits, green void diamonds, black stars and magenta crosses correspond respectively to the cases where one, two and four surface layers are allowed to be mobile.  }\label{figSus}
\end{figure}

The magnetic susceptibility and the heat capacity are shown in Fig. \ref{figSus} where the same observation is made: only when all layers are allowed to be mobile that the sublimation is a first-order transition.  Note that the small peaks at low $T$ correspond to the surface evaporation.



\begin{figure}[h!]
\centering
\includegraphics[scale=0.38]{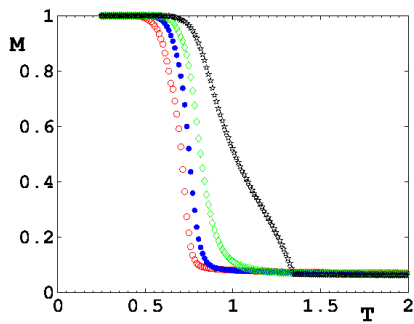}
\caption{(Color online) Layer magnetizations for the first four layers. Red void circles, blue filled circles, green diamonds and black stars are the magnetizations of the first, second, third and fourth layers. See text for comments. }\label{figSM}
\end{figure}

We show in Fig. \ref{figSM} the layer magnetization in the cases where one, two and four surface layers are mobile.  As seen, the layer next to the solid substrate is ``retained" by the latter up to the bulk transition occurring at $T_c\simeq 1.330$.  Other layers are evaporated starting from the first layer, at temperatures well below $T_c$.

%


To close this section let us compare the results obtained for two system shapes $20 \times 20 \times 40$ and $40 \times 20 \times 20$ with $c=50\%$.  It is obvious that the second shape has a larger free surface which facilitates the evaporation. As a consequence, there is no first-order transition because the solid core disappears at a temperature lower than the Potts transition temperature $T_c\simeq 1.330$ at the size $20 \times 20 \times 40$.
\begin{figure}[h!]
\centering
\includegraphics[scale=0.38]{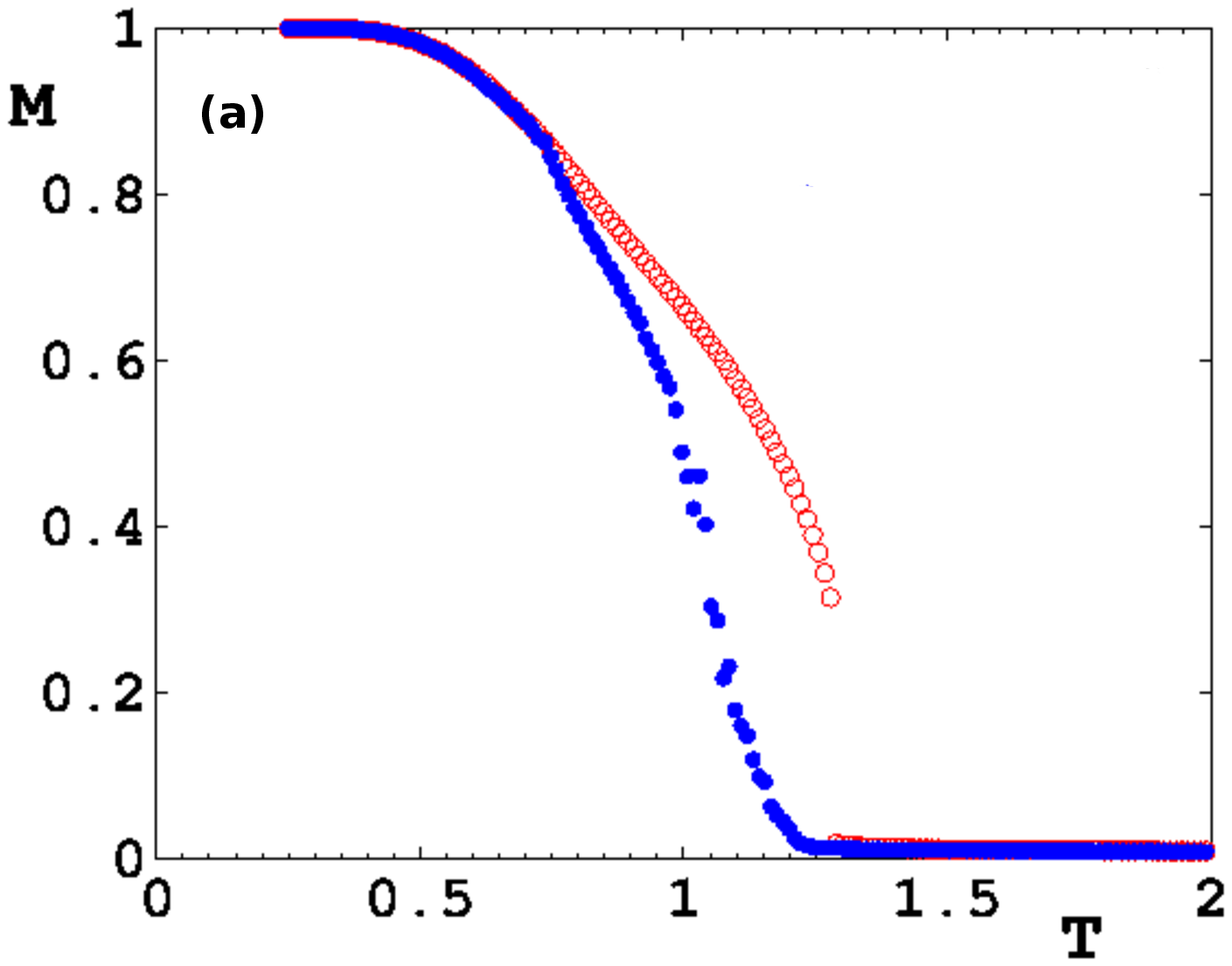}
\includegraphics[scale=0.38]{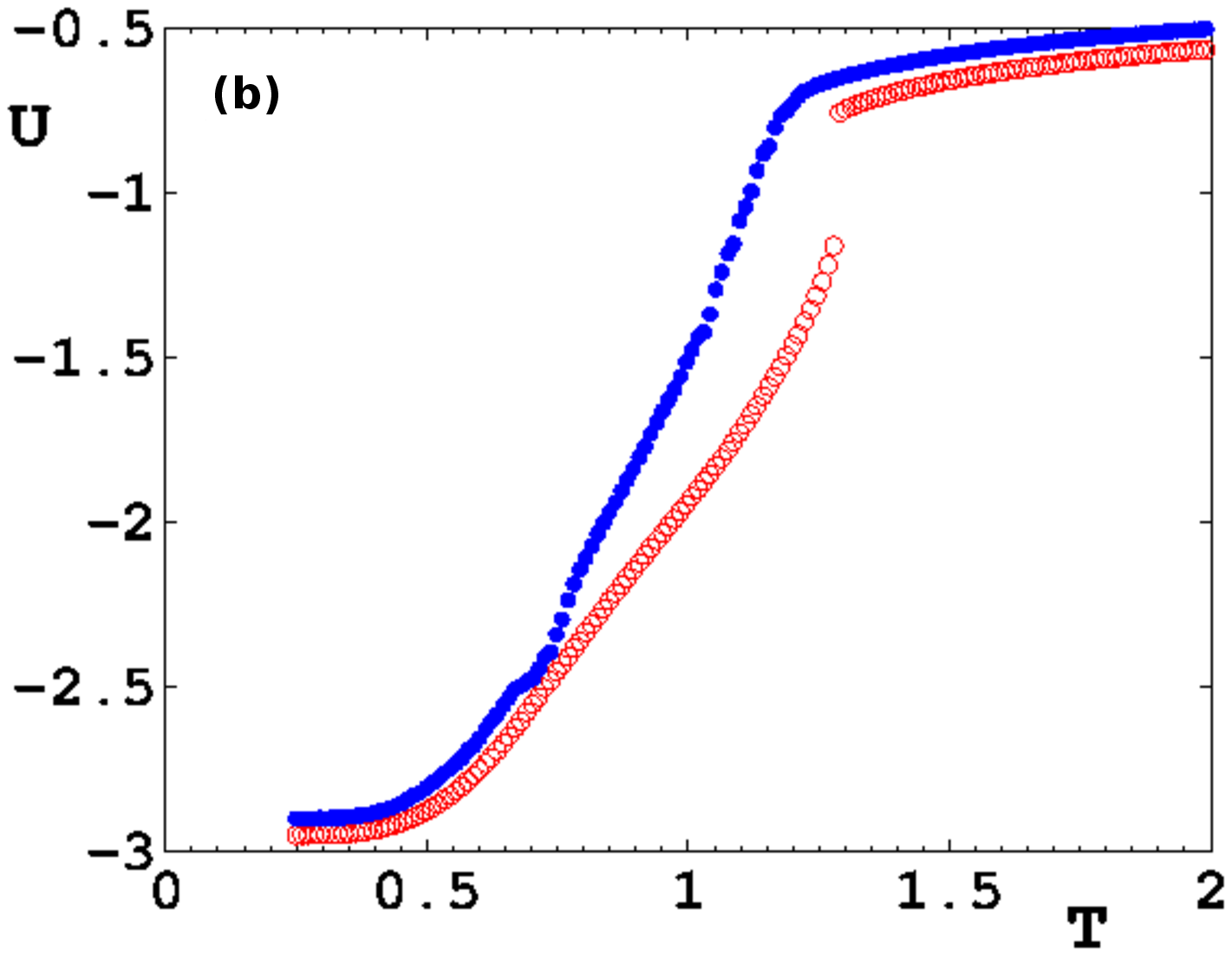}
\includegraphics[scale=0.38]{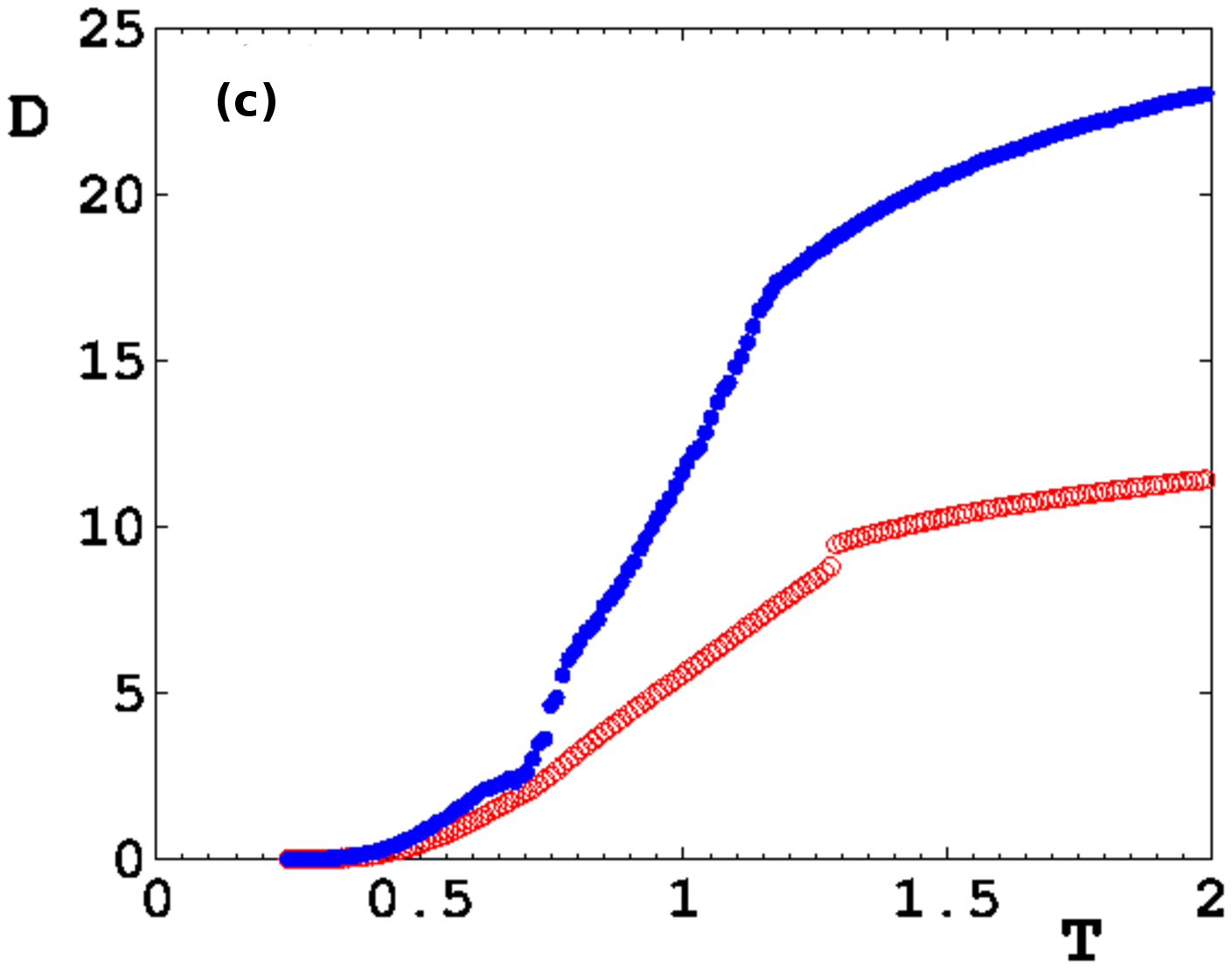}
\caption{(Color online) (a) Magnetization, (b) energy and (c) diffusion coefficient for two system shapes $20\times20\times40$ (red void circles) and $40\times20\times20$ (blue filled circles) at $c=50\%$. }\label{figShape}
\end{figure}

%

\section{Conclusion}\label{sectionConcl}

In this paper, we studied the properties of the mobile Potts model by the use of a mean-field theory and Monte Carlo simulations. The two methods confirm the first-order character of the phase transition in the bulk with $q=6$.
As discussed in the Introduction, the mean-field approach does not consider the real-time dynamics of the particles on the lattice sites. Rather, it considers the average numbers of particles per site.  In other words, it is equivalent to taking the spatial average first before considering the interaction between the particles uniformly distributed on lattice sites. Such a mean-field average is often used while dealing with disordered systems (dilution, bond-disorder, ...).  In MC simulations, the local environment of each particle is first taken into account before calculating its average over all particles. During the MC averaging, all local situations are expected to be taken into account in the final results.  Hence the mean-field approximation takes the spatial average before the ensemble average, while in MC simulations the calculation is first done for each spatial particle configuration and the statistical average is next made over configurations.  Furthermore the mean field approximation is applied to the diluted Potts model which is somewhat different than the mobile Potts model. In the MC simulations of the mobile Potts model a particle can be moved to a nearby vacant site while in the diluted Potts model a particle could be moved to a vacant site anywhere on the lattice.  This difference is expected to be important for the kinetics but not for the thermodynamics of the two models.

From a finite-size scaling we showed that the transition of an evaporating solid belongs to the $q=6$ localized Potts model. The reason is that a portion of the low-$T$ solid phase of the mobile Potts model  still remains solid at the transition temperature of the localized Potts model so that the orientational disordering of Potts spins occurs in this solid portion before the complete melting. Mean-field results for various parameters in the phase space are shown and discussed. In particular, we showed that there exists a threshold value of the chemical potential above which there is a solid-gas transition.  Monte Carlo simulations have been carried out to studied the surface evaporation behavior: we found that atoms are evaporated little by little from the surface at temperatures much lower than the bulk transition.
We believe that the model presented in this paper, though simple, possesses the essential evaporation properties.

\acknowledgments

ABR acknowledges financial support from CNRS.  MK is grateful for a visiting professorship's grant from the University of Cergy-Pontoise and for the warm hospitality extended to him during this working visit.


%



\end{document}